\documentclass[aps,prb,reprint]{revtex4-2}

\usepackage{graphicx} 		
\usepackage{amssymb}
\usepackage{amsmath}
\usepackage{dcolumn}		
\usepackage{bm}			
\usepackage{geometry} 
\begin{document}

\title{Temperature dependence of Bloch equations}

\author{David H. Gultekin}\email{david.gultekin@aya.yale.edu}

\affiliation{Department of Electrical Engineering, Yale University \\
Department of Diagnostic Radiology, Yale University}

\date{November 15, 2002}

\clearpage            

\begin{abstract}
We review the temperature dependence of Bloch equations. The
temperature dependence of nuclear magnetization in thermodynamic
equilibrium and relaxation are reviewed, and relations are
developed to account for the temperature effects in a nuclear
magnetic resonance experiment. It is found that the temperature
dependence of the Bloch equations precisely accounts for the
temperature effects observed in a nuclear magnetic resonance
experiment.

\end{abstract}

\clearpage

\maketitle

\section{\label{sec:level1}Introduction}

The temperature dependence of Bloch equations are derived to
account for the effects of heat on nuclear magnetization and
relaxation in an experiment. The temperature dependences of nuclear
magnetization in thermodynamic equilibrium and relaxation are
reviewed through Boltzmann's distribution and Bloch's relaxation
equations. The temperature effects are described simply by
the temperature dependence of Bloch equations.

\section{\label{sec:level2}Theory}

The Bloch equations~\cite{Bloch46} as functions of temperature for
nuclear magnetization relaxation can be written as

\begin{eqnarray}
\frac{\partial M_x(t,T)}{\partial t}&=& \gamma ({\bf M} \times
{\bf B})_x-\frac{M_x(t,T)}{T_2(T)}\label{B1}
\\
\frac{\partial M_y(t,T)}{\partial t}&=& \gamma ({\bf M}\times
{\bf B})_y-\frac{M_y(t,T)}{T_2(T)}\label{B2}
\\
\frac{\partial M_z(t,T)}{\partial t}&=& \gamma ({\bf M}\times
{\bf B})_z-\frac{M_z(t,T)-M(0,T)}{T_1(T)}\label{B3}
\end{eqnarray}

where $T_2(T)$ and $T_1(T)$ are transverse and longitudinal
magnetization relaxation times, respectively, and they are
functions of temperature~\cite{BPP48}.

In the rotating frame of reference following a short external
transverse field, Larmor precession terms~\cite{Larmor1897} drop, and Bloch equations reduce to

\begin{eqnarray}
\frac{\partial M_x(t,T)}{\partial t}&=&
-\frac{M_x(t,T)}{T_2(T)}\label{BR1}
\\
\frac{\partial M_y(t,T)}{\partial t}&=&
-\frac{M_y(t,T)}{T_2(T)}\label{BR2}
\\
\frac{\partial M_z(t,T)}{\partial t}&=&
-\frac{M_z(t,T)-M(0,T)}{T_1(T)}\label{BR3}
\end{eqnarray}

as functions of relaxation times and absolute temperature.
Combining Eqs.~(\ref{BR1}) and~(\ref{BR2}) in complex plane and
representing its magnitude as $M_{xy}(t,T)$, we have

\begin{eqnarray}
\frac{\partial M_{xy}(t,T)}{\partial t}&=&
-\frac{M_{xy}(t,T)}{T_2(T)}\label{BRT}
\\
\frac{\partial M_z(t,T)}{\partial t}&=&
-\frac{M_z(t,T)-M(0,T)}{T_1(T)}\label{BRL}
\end{eqnarray}

for the transverse and longitudinal components of magnetization
relaxation. We now review the temperature dependence of
equilibrium magnetization, transverse and longitudinal
magnetization relaxation.

\subsection{\label{sec:level2}Equilibrium Magnetization}

The nuclear magnetization in thermodynamic equilibrium is based on
Boltzmann's relation~\cite{Langevin1905} as

\begin{equation}\label{ch2m0T}
M(0,T)=\frac{N\gamma^2\hbar^2I(I+1)B_0}{3k_BT}
\end{equation}

where $\gamma$ is gyromagnetic ratio and $I$ is spin quantum
number. Substituting nuclear magnetic moment for spin quantum
number as $\mu^2=\gamma^2\hbar^2 I(I+1)$, we have

\begin{equation}\label{a1a}
M(0,T)=\frac{N\mu^2 B_0}{3k_BT}
\end{equation}

which can be written as

\begin{equation}\label{a1b}
M(0,T)=\chi(0,T) B_0
\end{equation}

where $\chi(0,T)$ is the static nuclear susceptibility. The
dependence of equilibrium magnetization on temperature is known as
Curie's~\cite{Curie1895} law. From~(\ref{a1a}), the temperature dependence of
equilibrium magnetization is then

\begin{equation}\label{a2}
\frac{\partial M(0,T)}{\partial T}=-\frac{M(0,T)}{T}
\end{equation}

and in terms of logarithmic derivatives equals to

\begin{equation}\label{a3}
\frac{\partial\ln M(0,T)}{\partial T}=-\frac{1}{T}
\end{equation}

as the negative of the reciprocal absolute temperature.

\subsection{\label{sec:level2}Transverse Magnetization}

The magnetization relaxation equation in the transverse plane

\begin{equation}\label{BT21}
\frac{\partial M_{xy}(t,T)}{\partial t}=-\frac{M_{xy}(t,T)}{T_2(T)}
\end{equation}

can be written in terms of logarithmic derivatives as

\begin{equation}\label{BT22}
\frac{\partial\ln M_{xy}(t,T)}{\partial t}=-\frac{1}{T_2(T)}
\end{equation}

Then, the first order derivative of Eq.~(\ref{BT22}) with respect
to temperature as

\begin{equation}\label{BT23}
\frac{\partial}{\partial T}\left[\frac{\partial\ln
M_{xy}(t,T)}{\partial t}\right] =\frac{1}{T_2^2(T)}\frac{\partial
T_2(T)}{\partial T}
\end{equation}

gives the dependence of the logarithm of transverse magnetization
on the relaxation time and absolute temperature.

Integration of Eq.~(\ref{BT22}) for a time interval as

\begin{equation}\label{T21}
\int_0^t\frac{dM_{xy}(t,T)}{M(t,T)}=-\int_0^t \frac{dt}{T_2(T)}
\end{equation}


gives the following solution for transverse magnetization

\begin{equation}\label{T23}
\ln\frac{M_{xy}(t,T)}{M(0,T)}=-\frac{t}{T_2(T)}
\end{equation}

and in exponential form as

\begin{equation}\label{T23a}
M_{xy}(t,T)=M(0,T)e^{-\frac{t}{T_2(T)}}
\end{equation}

where $M(0,T)$ is the magnetization in thermodynamic equilibrium.
A first order logarithmic derivative of Eq.~(\ref{T23a}) with
respect to temperature as

\begin{equation}\label{T24}
\frac{\partial\ln M_{xy}(t,T)}{\partial T} =-\frac{1}{T}+
\frac{t}{T_2^2(T)}\frac{\partial T_2(T)}{\partial T}
\end{equation}

gives the temperature dependence of the logarithm of magnetization in
terms of reciprocal temperature, and transverse relaxation time ${T_2(T)}$, as being linear in
relaxation time ${t}$. Now, a first order derivative of Eq.~(\ref{T24})
with respect to time as

\begin{equation}\label{BT25}
\frac{\partial}{\partial t}\left[\frac{\partial\ln
M_{xy}(t,T)}{\partial T}\right] = \frac{1}{T_2^2(T)}\frac{\partial
T_2(T)}{\partial T}
\end{equation}

simply gives the dependence of the logarithm of transverse
magnetization on the relaxation time and absolute
temperature~\cite{Gultekin02}. As it can be seen from
Eq.~(\ref{T24}), the temperature dependence of transverse
magnetization vanishes when $t=t_c$, the critical
relaxation time or echo time defined by

\begin{equation}\label{T26}
t_c=\frac{T_2^2(T)}{T}\left[\frac{\partial T_2(T)}{\partial
T}\right]^{-1}
\end{equation}

as verified experimentally for a number of
substances~\cite{Gultekin02} using a spin echo pulse
sequence~\cite{Hahn50,CarrPurcell54}.

Taking the first order derivative of Eq.~(\ref{BT21}) with respect
to temperature as

\begin{widetext}
\begin{eqnarray}\label{T27}
\frac{\partial^2M_{xy}(t,T)}{\partial T \partial t}=
\left[-\frac{1}{T_2(T)}\frac{\partial\ln M_{xy}(t,T)}{\partial T}
+\frac{1}{T_2^2(T)}\frac{\partial T_2(T)}{\partial T} \right]
M_{xy}(t,T)
\end{eqnarray}
\end{widetext}

and substituting Eq.~(\ref{T24}) into Eq.~(\ref{T27}), we find

\begin{widetext}
\begin{eqnarray}\label{T28}
\frac{\partial^2M_{xy}(t,T)}{\partial T \partial t}=
\left[\frac{1}{TT_2(T)}+\frac{1}{T_2^2(T)}\frac{\partial
T_2(T)}{\partial T} -\frac{t}{T_2^3(T)}\frac{\partial
T_2(T)}{\partial T} \right] M_{xy}(t,T)
\end{eqnarray}
\end{widetext}

the temperature dependence of Bloch's equation for transverse
relaxation.

\subsection{\label{sec:level2}Longitudinal Magnetization}

The Bloch's equation for longitudinal magnetization relaxation

\begin{equation}\label{BT11}
\frac{\partial M_z(t,T)}{\partial t}
=-\frac{\left[M_z(t,T)-M(0,T)\right]}{T_1(T)}
\end{equation}

can be written in its logarithmic derivative form as

\begin{equation}\label{BT12}
\frac{\partial\ln\left[M_z(t,T)-M(0,T)\right] }{\partial t}
=-\frac{1}{T_1(T)}
\end{equation}

Taking the first order derivative of Eq.~(\ref{BT12}) with respect
to temperature as

\begin{equation}\label{BT13}
\frac{\partial}{\partial T}\left[
\frac{\partial\ln\left[M_z(t,T)-M(0,T)\right] }{\partial t}\right]
=\frac{1}{T_1^2(T)}\frac{\partial T_1(T)}{\partial T}
\end{equation}

gives the dependence of the logarithm of the differential
longitudinal magnetization on the recovery time and absolute
temperature.

Integration of Eq.~(\ref{BT11}) for a time interval as

\begin{equation}\label{T11}
\int_0^t\frac{dM_z(t,T)}{\left[M_z(t,T)-M(0,T)\right]} =
-\int_0^t\frac{dt}{T_1(T)}
\end{equation}

results in the following solution for the differential
longitudinal magnetization

\begin{equation}\label{T13}
\ln\frac{\left[M_z(t,T)-M(0,T)\right]}{\left[M_z(0,T)-M(0,T)\right]}=
-\frac{t}{T_1(T)}
\end{equation}

Using a flip angle of $\theta=\pi/2$, we have $M_z(0,T)=0$ and

\begin{equation}\label{T14}
\left[M_z(t,T)-M(0,T)\right]=-M(0,T)e^{-\frac{t}{T_1(T)}}
\end{equation}

for the magnetization recovery, and using a flip angle of
$\theta=\pi$, we have $M_z(0,T)=-M(0,T)$ and

\begin{equation}\label{T15}
\left[M_z(t,T)-M(0,T)\right]=-2M(0,T)e^{-\frac{t}{T_1(T)}}
\end{equation}

for the magnetization inversion recovery.

A first order logarithmic derivative of Eq.~(\ref{T14}) or
Eq.~(\ref{T15}) with respect to temperature as

\begin{equation}\label{T16}
\frac{\partial\ln\left[M_z(t,T)-M(0,T)\right]}{\partial T}
=-\frac{1}{T}+\frac{t}{T_1^2(T)}\frac{\partial T_1(T)}{\partial T}
\end{equation}

gives the temperature dependence of the differential longitudinal
magnetization. And a first order derivative of Eq.~(\ref{T16})
with respect to time as

\begin{equation}\label{T17}
\frac{\partial}{\partial t}\left[
\frac{\partial\ln\left[M_z(t,T)-M(0,T)\right] }{\partial T}\right]
=\frac{1}{T_1^2(T)}\frac{\partial T_1(T)}{\partial T}
\end{equation}

gives the dependence of the logarithm of the differential
longitudinal magnetization on the recovery time and absolute
temperature.

Now,  taking the first order derivative of Eq.~(\ref{BT11}) with
respect to temperature

\begin{widetext}
\begin{eqnarray}\label{BT11T}
\frac{\partial^2M_z(t,T)}{\partial T\partial
t}=-\frac{1}{T_1(T)}\frac{\partial}{\partial
T}\left[M_z(t,T)-M(0,T)\right]
 +\frac{1}{T_1^2(T)}\frac{\partial
T_1(T)}{\partial T}\left[M_z(t,T)-M(0,T)\right]
\end{eqnarray}
\end{widetext}

and substituting the temperature dependence of Eq.~(\ref{T14})
into Eq.~(\ref{BT11T}) we find

\begin{widetext}
\begin{eqnarray}\label{BT11Ta}
\frac{\partial^2M_z(t,T)}{\partial T\partial t}=
\left[\frac{1}{TT_1(T)}
 +\frac{1}{T_1^2(T)}\frac{\partial
T_1(T)}{\partial T}
 -\frac{t}{T_1^3(T)}\frac{\partial
T_1(T)}{\partial T}\right]
\left[M_z(t,T)-M(0,T)\right]
\end{eqnarray}
\end{widetext}

for the temperature dependence of Bloch's equation in longitudinal
relaxation.

Alternatively, writing the Eq.~(\ref{T14}) as

\begin{equation}\label{T18}
M_z(t,T)=M(0,T)(1-e^{-\frac{t}{T_1(T)}})
\end{equation}

and taking its first order logarithmic derivative with respect to
temperature

\begin{equation}\label{T19}
\frac{\partial\ln M_z(t,T)}{\partial T}
=-\frac{1}{T}-\frac{t}{T_1^2(T)}\frac{\partial T_1(T)}{\partial T}
\frac{e^{-\frac{t}{T_1(T)}}}{1-e^{-\frac{t}{T_1(T)}}}
\end{equation}

and expanding the exponential term in Taylor series and taking
only the first terms as

\begin{equation}\label{T110}
\frac{e^{-\frac{t}{T_1(T)}}}{1-e^{-\frac{t}{T_1(T)}}}=
\frac{1-\frac{t}{T_1(T)}-...}{1-1+\frac{t}{T_1(T)}+...}=
\frac{T_1(T)}{t}-1
\end{equation}

and substituting the result into Eq.~(\ref{T19}), the temperature
dependence of the logarithm of longitudinal magnetization becomes

\begin{equation}\label{T111}
\frac{\partial\ln M_z(t,T)}{\partial T}
=-\frac{1}{T}-\frac{1}{T_1(T)}\frac{\partial T_1(T)}{\partial T}
+\frac{t}{T_1^2(T)}\frac{\partial T_1(T)}{\partial T}
\end{equation}

Now, a first order temporal derivative of Eq.~(\ref{T111}) as

\begin{equation}\label{T112}
\frac{\partial}{\partial t}\left[\frac{\partial\ln
M_z(t,T)}{\partial T}\right] = \frac{1}{T_1^2(T)}\frac{\partial
T_1(T)}{\partial T}
\end{equation}

simply gives the dependence of the logarithm of the longitudinal
magnetization on the recovery time and absolute temperature.

\section{Summary}

The temperature dependence of Bloch's equations are

\begin{widetext}
\begin{eqnarray}
\frac{\partial^2 M_{xy}(t,T)}{\partial T \partial t} &=&
\left[\frac{1}{TT_2(T)} +\frac{1}{T_2^2(T)}\frac{\partial
T_2(T)}{\partial T}-\frac{t}{T_2^3(T)}\frac{\partial
T_2(T)}{\partial T}\right]M_{xy}(t,T)
\\
\frac{\partial^2 M_z(t,T)}{\partial T \partial t} &=&
\left[\frac{1}{TT_1(T)} +\frac{1}{T_1^2(T)}\frac{\partial
T_1(T)}{\partial T}-\frac{t}{T_1^3(T)}\frac{\partial
T_1(T)}{\partial T}\right]\left[M_z(t,T)-M(0,T)\right]
\end{eqnarray}
\end{widetext}

for the transverse and longitudinal magnetization relaxation.

Finally, the temperature dependence of Bloch's equations in terms
of logarithmic derivatives are

\begin{widetext}
\begin{eqnarray}
\frac{\partial}{\partial T}\left[\frac{\partial\ln
M_{xy}(t,T)}{\partial t}\right] &=& \frac{1}{T_2^2(T)}\frac{\partial
T_2(T)}{\partial T}\label{BRT1}
\\
\frac{\partial}{\partial T}\left[\frac{\partial\ln
M_z(t,T)-M(0,T)}{\partial t}\right] &=&
\frac{1}{T_1^2(T)}\frac{\partial T_1(T)}{\partial T}\label{BRT2}
\end{eqnarray}
\end{widetext}

for the transverse and longitudinal magnetization relaxation. The
temperature is taken as not varying or very slowly varying with
time compared to the magnetization and relaxation processes.

\section*{Acknowledgments} 
We wish to acknowledge the support for this
research from the Department of Diagnostic Radiology at Yale University.

\appendix

\section{Temperature Effects on Nuclear Shielding}\label{AB}

The equilibrium magnetization is proportional to the magnetic field as

\begin{equation}\label{b1}
M(0,T)=\chi(0,T)B
\end{equation}

where $B$ is the magnetic field at the nuclei of the atoms, and it
is proportional to the static magnetic field strength through the
nuclear shielding. The temperature effect can be taken into
account as

\begin{equation}\label{b2}
M(0,T)=\chi(0,T)(1-\sigma(T))B_0
\end{equation}

where $\sigma(T)$ is nuclear shielding and it is temperature
dependent~\cite{RamseyShield}. The temperature dependence of
equilibrium magnetization is then

\begin{equation}\label{b3}
\frac{1}{M(0,T)}\frac{\partial M(0,T)}{\partial
T}=-\frac{1}{T}-\frac{1}{1-\sigma(T)}
\frac{\partial\sigma(T)}{\partial T}
\end{equation}

in terms of the temperature dependence of nuclear shielding and
absolute temperature. While the effects of temperature dependent
nuclear shielding on precession are significant in terms of the
nuclear spin phase shift, its effects on the magnitude of nuclear
magnetization are small as

\begin{equation}\label{b4}
\left|\sigma(T)\right|\ll 1, \hspace{.5in}
\left|\frac{\partial\sigma(T)}{\partial T}\right| \ll 1
\end{equation}

and therefore the temperature dependence of equilibrium
magnetization will reduce to

\begin{equation}\label{b5}
\frac{1}{M(0,T)}\frac{\partial M(0,T)}{\partial T}=-\frac{1}{T}
\end{equation}

or in logarithmic derivatives to

\begin{equation}\label{b6}
\frac{\partial\ln M(0,T)}{\partial T}=-\frac{1}{T}
\end{equation}

as given earlier.

\bibliography{TmagnetPR}

\end{document}